\documentclass[twocolumn]{elsart3p}
\usepackage{graphicx,amssymb,epsfig,url}

\def\Cpp{C$^{++}$}

\begin{document}

\begin{frontmatter}

\title{ACTIVIA: Calculation of isotope production cross-sections and yields}
\author{J.~J.~Back\corauthref{cor}},
\corauth[cor]{Corresponding author.}
\ead{J.J.Back@warwick.ac.uk}
\author{Y.~A.~Ramachers}
\ead{Y.A.Ramachers@warwick.ac.uk}

\address{Department of Physics, University of Warwick,
Coventry, CV4 7AL, United Kingdom}

\date{15 January 2008}

\begin{abstract}
We present a C$^{++}$ computer package, ACTIVIA, that can calculate target-product 
cross-sections and the production and decay yields of isotopes from cosmic ray activation
using data tables and semi-empirical formulae. We describe the structure and 
user interface of the computer code as well as provide comparisons between the calculations and 
experimental results. We also outline suggestions on how the code can be improved and 
extended for other applications.
\end{abstract}

\end{frontmatter}

\section{Introduction}
\label{sec:Introduction}

There are many observations made in astrophysics, astronomy
and particle physics that require knowledge of the cross-sections of nuclear 
processes at high energies. The physics of cosmic rays is one example.
Understanding how they interact with matter allows us to study
the composition of the interstellar medium, the Galaxy and the solar system.
Estimating the cosmic ray background radiation in spacecraft and isotope
production in accelerator facilities
are other important applications. For the case of low background experiments, it is 
necessary to know which isotopes are formed by cosmic
ray activation inside target materials before they are shielded underground.
Any long-lived isotopes must be 
accounted for in the signal data analysis, and this requires information about 
cross-sections and production rates. There are measurements available for use
in such studies, but normally only for a restricted energy range which does not
cover the total energy spectrum of the input beam. Some important reactions
for low background experiments, such as $^{60}$Co produced from natural tellurium by cosmic
ray activation, have essentially no experimental data available. Accordingly, a set 
of semi-empirical formulae have been developed by Silberberg and Tsao~\cite{ST1,ST2} to 
estimate the cross-sections of various nuclear processes, such as spallation and fission, 
which can be used to calculate the production yields of nuclear isotopes from cosmic ray activation.
The parameters in the formulae, assumed to be the same for proton and neutron beams, 
have been tuned to best fit the available experimental results.

In this paper we describe a new \Cpp\ package, ACTIVIA, that uses a combination
of semi-empirical formulae and tables of data based on experimental results
to calculate the cross-section, production rates
and yields of radioactive isotopes from cosmic ray activation.
Section~\ref{sec:formulae} describes the Silberberg-Tsao
semi-empirical formulae used for the cross-section calculations while Section~\ref{sec:datatables} 
explains how cross-sections can be evaluated if tables of data (in ASCII text format) are available.
Section~\ref{sec:decay} provides details on how the radioactive decay yields are calculated,
Section~\ref{sec:code} explains the basic structure of the code, while Section~\ref{sec:comparisons} 
shows comparisons between the calculations and experimental data.
Finally, in Section~\ref{sec:extensions} we describe possible 
modifications and extensions that could be made to the code if other 
input beam spectra or nuclear physics processes are required.

\section{Semi-empirical formulae}
\label{sec:formulae}

We use the extensive set of semi-empirical formulae from Silberberg and 
Tsao~\cite{ST1,ST2}
to calculate cross-sections for isotope production when a proton or neutron beam 
(e.g.\,cosmic rays) hits a given target. The general form of the cross-section equation 
for a product $(Z,A)$ from a target $(Z_t,A_t)$ is
\begin{eqnarray}
\label{eqn:STSigma}
\sigma & = & \sigma_0 f(A) f(E) e^{-P\Delta A} \times \\ \nonumber
& & {\rm{exp}}\left(-R|Z - SA + TA^2 + UA^3|^{\nu}\right) \Omega \eta \xi,
\end{eqnarray}
where $E$ is the beam energy (in units of MeV), $\Delta A$ is the difference between the 
target and product mass numbers ($A_t - A$), while $f(A)$ and $f(E)$ are correction
factors usually applied for heavy targets ($A_t > 30$) and when $\Delta A$ is 
large ($\gtrsim 10$).
The normalisation factor $\sigma_0$ is the cross-section (in millibarns)
at the energy $E_0$ ($\sim 3$\,GeV), above which cross-sections are assumed to be
independent of energy.
The first exponential term describes the reduction of the cross-section as $\Delta A$ 
increases, where $P$ is a function of target mass $A_t$ and $E$. The second 
exponential term describes the distribution of cross-sections for several isotopes 
for a given element  of atomic number $Z$; its parameters ($R$, $S$, $T$, $U$ and $\nu$) are, 
in general, functions of target and product $Z$ and $A$ values as well as energy. 
The parameter $R$ represents the width of the cross-section distribution amongst 
the isotopes, while $S$ describes the peaks for various mass  numbers $A$. 
The parameter $T$ describes the increase in the number of neutrons as $A$ increases, 
while $U$ is equal to $3.0\times 10^{-7}$ and represents a small correction to improve 
the overall agreement with available experimental data. The power exponent $\nu$ varies 
from 1.3 to 2.0 depending on the target and product isotopes. The parameter $\Omega$ 
represents a nuclear structure factor, $\eta$ is a nuclear pairing factor (for even 
and odd $Z$ and neutron ($N$) numbers), while $\xi$ is an enhancement factor for 
light evaporation products; these parameters are of order unity.
We assume that Eq.~\ref{eqn:STSigma} is applicable to both proton and neutron beams,
with identical formulae and parameters.

\begin{figure}[!htb]
\begin{center}
\includegraphics*[angle=0,width=0.5\textwidth]{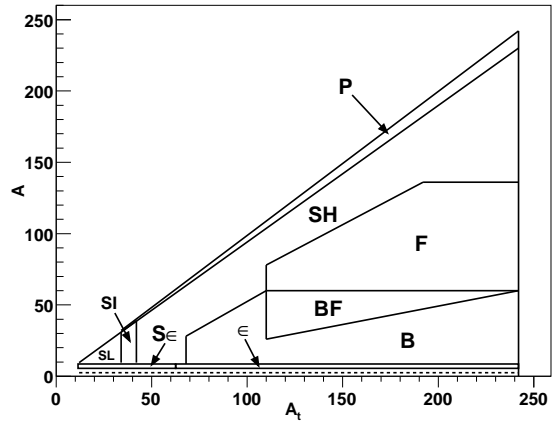}
\caption
{Domains of applicability of different parameters of Eq.~\ref{eqn:STSigma} as a function
of target mass $A_t$ and product mass $A$, which are implemented in the code.
P denotes the reactions that are mainly peripheral; SL, SI
and SH denote the spallation of light, intermediate and heavy targets, respectively. The fission
region (including contributions from spallation reactions) is denoted by F, the symbol B represents
the breakup region in which light nuclei are evaporated, while BF represents the combination of
fission and breakup reactions. The evaporation of light nuclei occurs in region $\epsilon$, with
S$\epsilon$ combining evaporation and spallation processes. These are the
same regions as those in Ref.~\cite{ST1}. Finally, the dotted line represents
the tritium production region~\cite{tritium}.}
\label{fig:STRegions}
\end{center}
\end{figure}

Figure~\ref{fig:STRegions} shows the various regions for different hadronic processes which 
can be calculated using Eq.~\ref{eqn:STSigma} and which are implemented in the computer code.
See Silberberg and Tsao~\cite{ST1,ST2} for detailed information about the 
parameterisations of the various factors in Eq.~\ref{eqn:STSigma}.
The code also calculates the cross-sections for tritium 
production using the formulae from Konobeyev and Korovin~\cite{tritium}.

%

\section{Data tables}
\label{sec:datatables}

There are extensive sets of nuclear data 
(e.g.\, the EXFOR database~\cite{EXFOR}) that can be used to provide
cross-sections as a function of energy for target-product isotope pairs. If present, 
these (ASCII text) tables
are used instead of the semi-empirical formulae. A linear interpolation is performed between
data points to provide the cross-section (in mb) at a given beam energy (in MeV). 
It is important that the data be in the correct units, have enough energy bins to be useful 
and have a contiguous, increasing energy range. Equation~\ref{eqn:STSigma} is used to 
calculate the cross-section if any data value is below a selectable lower limit
(recommended to be set at 0.001\,mb).
If two or more separate energy regions are needed, then the data from these must be combined 
into the same table. For the case of large energy differences in between neighbouring regions, 
it is sensible to set the cross-sections to zero just above and below the range limits in 
the table so that the semi-empirical formulae are used instead.
Otherwise, the linear interpolation would only give a weighted average of the cross-sections
between the data limits of neighbouring energy regions, which may not be reliable enough.
The code can use MENDL-2P tables~\cite{MENDL2P} for intermediate energy
(up to 200\,MeV) cross-sections.

\section{Radioactive decay yields}
\label{sec:decay}

\begin{figure}[!thb]
\begin{center}
\includegraphics*[angle=0,width=0.5\textwidth]{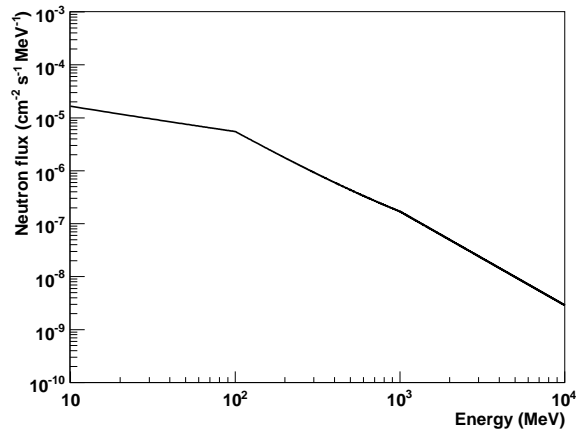}
\caption{Energy spectrum for cosmic ray neutrons at sea 
level based on the parameterisation from Armstrong~\cite{Armstrong} and Gehrels~\cite{Gehrels}.}
\label{fig:cosrayspectrum}
\end{center}
\end{figure}

Consider a target comprised of several isotopes with the same atomic number $Z_t$ but
having different mass numbers $A_{ti}$ each with a relative abundance fraction $f_i$ such that 
$\sum_i f_i = 1$. This target is exposed to a beam with an energy spectrum $d\phi/dE$ 
(cm$^{-2}$\,s$^{-1}$\,MeV$^{-1}$), such as cosmic ray neutrons shown in
Fig.~\ref{fig:cosrayspectrum}, which is based on the parameterisation from 
Armstrong~\cite{Armstrong} and Gehrels~\cite{Gehrels}.
The production rate (kg$^{-1}$\,day$^{-1}$) of an isotope
$j$ will be
\begin{equation}
\label{eqn:yield1}
Y_j = C \sum_i {\frac{f_i}{A_{ti}} \int \frac{d\phi}{dE} \sigma_{ij}(E) dE },
\end{equation}
where $C$ is a normalisation factor and $\sigma_{ij}(E)$ is the cross-section (mb) for the
target-product isotope pair at energy $E$ (MeV).
The integral is over the required energy range of the input beam spectrum, while
the sum is over all of the target isotopes. Here,
\begin{equation}
C = N_A \times 24 \times 3600 \times 10^{-24},
\end{equation}
where $N_A$ is Avogadro's number ($6.022 \times 10^{23}$\,mol$^{-1}$), the next
two numbers convert the production rate from per second to per day, while
the last factor accounts for the different units of area used for the flux (cm$^{2}$)
and cross-section (mb) quantities, as well as the unit of mass used for the atomic number $A_{ti}$
(converting g to kg).
If the target is exposed to the (continuous) beam for a time $t_{\rm{exp}}$ (days), then the
yield rate (kg$^{-1}$\,day$^{-1}$) of the radioactive product isotope $j$ with
a half-life of $t_{1/2}$ (days) is given by
\begin{equation}
\label{eqn:yield2}
Y_j^{\rm{exp}} = Y_j \left( 1 - e^{-\lambda_j t_{\rm{exp}}} \right),
\end{equation}
where $\lambda_j = {\rm{ln}}2/t_{1/2}$ and assuming no chained decays. 
If the target is removed from the beam and the product isotope 
is allowed to decay for a time $t_{\rm{dec}}$ (days), then the final product yield rate is
\begin{equation}
\label{eqn:yield3}
Y_j^{\rm{dec}} = Y_j^{\rm{exp}} e^{-\lambda_j t_{\rm{dec}}} = 
Y_j \left( 1 - e^{-\lambda_j t_{\rm{exp}}} \right) e^{-\lambda_j t_{\rm{dec}}}.
\end{equation}
Here, we only consider product isotopes with half-lives greater than one day. However, there
are many short-lived isotopes which can decay to these, and they are included in the
cross-section and yield calculations as ``side-branch nuclei''. For example, consider the decay
chain to the long-lived isotope $^{14}_{6}$C ($t_{1/2}$ = 5730\,years):
$^{14}_{4}{\rm{Be}} \rightarrow$ $^{14}_{5}$B $+ e^- + {\bar{\nu}}_e$ ($t_{1/2}$ = 4\,ms) and 
$^{14}_{5}{\rm{B}}  \rightarrow$ $^{14}_{6}$C $+ e^- + {\bar{\nu}}_e$ ($t_{1/2}$ = 14\,ms).
This means that both $^{14}_{4}$Be and $^{14}_{5}$B are considered side-branches of
$^{14}_{6}$C, and their cross-sections and yields will contribute to those for $^{14}_{6}$C.

\section{Code structure}
\label{sec:code}

The code, written in \Cpp\ to take advantage of object-oriented design principles,
is split into two main sections; the first part performs the calculation of
the cross-sections (Eq.~\ref{eqn:STSigma}) and production rates (Eq.~\ref{eqn:yield1}),
while the second part calculates the final yields of the radioactive isotope 
products (Eq.~\ref{eqn:yield3}). The basic structure of the code, designed to work in the 
Unix environment, is described below. Details on how to build and run
the code are found in the README file of the computer package~\cite{Code}.

\subsection{Input}

\begin{figure*}[!htb]
\begin{center}
\includegraphics[angle=-90,width=0.85\textwidth]{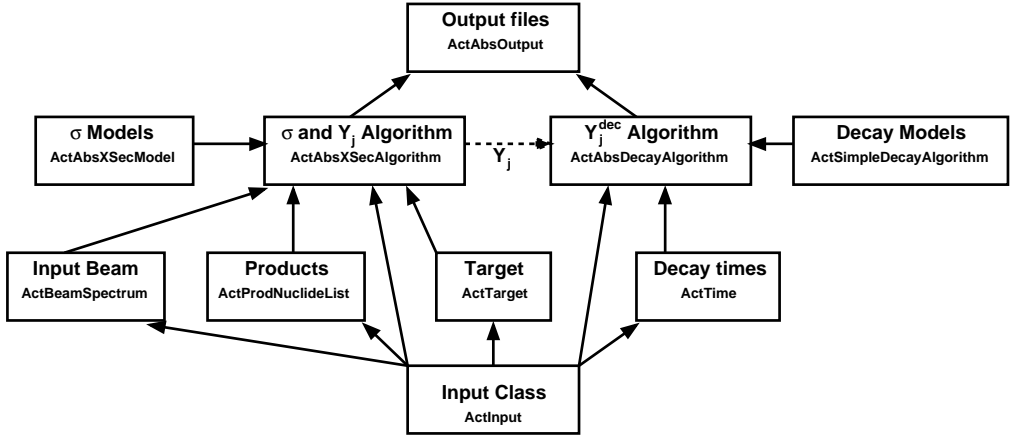}
\caption
{Diagram showing a simplified version of the program flow and \Cpp\ class structure.
The input specifies the
beam spectrum, the list of product and target isotopes, the decay times, as well
as algorithms for the cross-section and yield calculations. These algorithms use
the various inputs with the required cross-section and decay yield models to generate
the output files for further analysis.}
\label{fig:ProgramFlow}
\end{center}
\end{figure*}

There are several inputs that are necessary for the cross-section and yield
calculations. They are the target isotopes, a list of all allowed product
isotopes, the input beam spectrum, as well as the beam exposure and
target cooling times. These are specified by the ActAbsInput abstract interface, which
is a base class that contains several pure virtual functions
that must be implemented by any derived classes, such as ActInput. Figure~\ref{fig:ProgramFlow}
shows a simplified version of the overall program flow.

The class ActTarget defines the target element, together with its isotopes and their relative
abundance fractions $f_i$ (such that $\sum_i f_i = 1$). 
If a target is comprised of many elements (each with their own isotopes),
then the code must be run as many times as is necessary to get the results for each target element.
The results must then be combined at the end by the user.

The list of allowed products is defined in the ActProdNuclideList class,
which is built from an ASCII text file (``decayData.dat'') containing a complete list of radioactive 
isotopes found in the periodic table with half-lives equal to or greater than one day.
All target and product isotopes are represented by ActNuclide objects, which store
the atomic ($Z$) and mass ($A$) numbers as well as the half-life $t_{1/2}$. 
Product nuclide objects also contain the $Z$ and $A$ values of any side-branch nuclei
(defined at the end of Section~\ref{sec:decay}). Each nuclide is created
only once by the factory class ActNuclideFactory, independent of how many times it is
used in the code.

The input beam spectrum is defined through the abstract class ActBeamSpectrum, in which
the flux ($d\phi/dE$ in units of cm$^{-2}$\,s$^{-1}$\,MeV$^{-1}$) must be implemented 
by any derived class. At present, the only available input beam is
the cosmic ray spectrum shown in Fig.~\ref{fig:cosrayspectrum}, which is assumed to be valid
for neutrons and protons.
The ActBeamSpectrum class also stores the energy range and bin
width $\Delta E$, which are used to calculate the integral in Eq.~\ref{eqn:yield1}, 
as well as the atomic number
$Z_{\rm{beam}}$ and mass number $A_{\rm{beam}}$ of the beam particles, both equal to unity
for cosmic ray protons.

The final required input variables are the time $t_{\rm{exp}}$ that the target is exposed to the beam
and the time elapsed after the target is removed from the beam (decay or cooling time $t_{\rm{dec}}$), 
both in units of days. These are stored within the simple ActTime class.

The set-up of the calculations is managed by the ActInput class, as mentioned earlier, which
prompts the user to provide the required inputs using the standard \Cpp\ input streamer ``cin''.
Examples are provided in the README file of the package~\cite{Code}.

The algorithms for calculating the cross-sections and yields must also be selected in the input class.
At present, the cross-sections are calculated using Eq.~\ref{eqn:STSigma} (with any available data
tables as described in Section~\ref{sec:datatables}) by the ActSTXSecAlgorithm class, while the 
yields are evaluated using Eq.~\ref{eqn:yield3} by the ActSimpleDecayAlgorithm class.
We now describe how these algorithms work.

\subsection{Cross-sections}

The calculation of the cross-sections between all target and product isotope pairs is
controlled by the ActIsotopeProduction class, which retrieves information about the target,
the list of product isotopes, and the beam energy spectrum from the input class.

The cross-sections for each target isotope $(Z_t, A_t)$ are calculated using the ActProdXSecData
utility class. This loops over all available product isotopes $(Z,A)$ and calculates the cross-section
for each energy bin in the input beam spectrum. The values of $Z$ and $A$ must satisfy
the requirements $Z \leq Z_t + Z_{\rm{beam}}$ and $A \leq A_{t} + A_{\rm{beam}} - 1$, which
ensure that there are at least two isotope products in each reaction, i.e. the cross-sections
for any $^{A}_{Z}X + p \rightarrow ^{A+1}_{Z+1}Y + \gamma$ reactions are not calculated.
Any side-branch nuclei are included in the cross-section calculations. 
The energy of the reaction must be larger than
a threshold energy, typically of order 10\,MeV, defined as the mass excess 
from the target-product reaction, and is calculated using the semi-empirical formula from 
Seeger~\cite{Seeger}. Otherwise, the cross-section is set to zero and the next energy 
bin is considered. This requirement avoids calculating the cross-section of resonances for 
low-energy ($E \lesssim 10$\,MeV) reactions, which is beyond the scope of this work.

The required cross-section is retrieved from the ActSTXSecAlgorithm class.
This first checks if there are any cross-section data for the target-product isotope pair
at the given energy (above threshold). If no such data exists, or the 
cross-section is too small ($<0.001$\,mb, for example), then Eq.~\ref{eqn:STSigma} 
is used to calculate the cross-section. The algorithm selects which
calculation region in Fig.~\ref{fig:STRegions} is appropriate, i.e. what
parameters are needed for Eq.~\ref{eqn:STSigma}, and calls the class that
implements the specific Silberberg-Tsao model, such as spallation, fission or peripheral
reactions. The total cross-section is then just the sum of $\sigma(E)$ over all energy bins.
The calculations do not distinguish between ground or metastable states for the product nuclei.

\subsection{Yields}

The production rates of the isotopes (Eq.~\ref{eqn:yield1}) are calculated by ActIsotopeProduction 
when the cross-sections are evaluated. Following Eq.~\ref{eqn:yield1}, 
the production rate at the energy $E_k$ is given by
\begin{equation}
y_j(E_k) = C \sum_i \frac{f_i}{A_{ti}} \frac{d\phi}{dE} \sigma_{ij}(E_k) \Delta E,
\end{equation}
where $\Delta E$ is the energy bin width from the input beam spectrum.
The total production rate is given by the sum of $y_j(E_k)$ over all energy bins
\begin{equation}
Y_j = C \sum_i \frac{f_i}{A_{ti}} 
\left[ \sum_k \frac{d\phi}{dE} \sigma_{ij}(E_k) \Delta E \right],
\end{equation}
where the sum over $k$ specifies the required energy bin range. 
The ActSimpleDecayAlgorithm class implements Eq.~\ref{eqn:yield3} to find the 
final yield $Y^{\rm{dec}}_j$ for 
each product isotope $j$ from the given exposure and decay times.
This simple calculation ignores any chained decays, in which isotopes
can decay to several other long-lived nuclei, or when isotopes can be made from 
multiple decays. However, short-lived isotopes are included in the final product 
yields through the use of the side-branch nuclei, as mentioned in Section~\ref{sec:decay}.

\subsection{Output}

The results of the cross-section and yield calculations are stored in output files 
using the abstract interface ActAbsOutput, which contains several 
pure virtual functions that must be implemented by any derived class, such as 
ActStreamOutput for ASCII text output or ActROOTOutput for ROOT~\cite{ROOT} output.
These classes must be able to deal with three different formats: lines
of text (character strings or ActString objects), tables of numbers (ActOutputTable objects) 
and graphs (ActAbsGraph objects). The cross-section and yield algorithms are not concerned with
specific formats of output files; they only create tables and graphs which are then
passed onto the output class, which decides how to write out the information in whatever format
is required for further analysis.

The ActOutputTables store rows of numbers, where each column represents
a specific variable with its own unique label. Graphs are made up of a series of points, which are
represented by ActGraphPoint objects that can store several $y$ values for a given $x$ co-ordinate,
each with their own labels and units.
For example, ActXSecGraph has $x$ equal to the energy $E$ (MeV),
$y_1$ equal to the cross-section value (mb) and $y_2$ equal to the production rate
$Y_j$ (kg$^{-1}$ day$^{-1}$). ActDecayGraphs contain points storing the elapsed time (in days) 
and the corresponding isotope yield: ($x$, $y$) = ($t_{\rm{exp}	}$, $Y^{\rm{exp}}_j$) or 
($t_{\rm{exp}}+t_{\rm{dec}}$, $Y^{\rm{dec}}_j$).

Each output class must be able to deal with two output files; the first one contains
details from the cross-section and production rate calculations, while
the second one contains the yields from the isotope decay algorithm.

The total cross-section (mb) and production rate (kg$^{-1}$ day$^{-1}$) 
are stored in the cross-section output file. Also stored is the cross-section and production rate 
as a function of input beam energy (MeV), but only if 
a flag controlling the level of output detail is set appropriately. Again, the output class
must determine what to do when the level of detail is changed; the cross-section and yield
calculations are not affected by this. Finally, a summary table of
the yields of each product isotope at the start and end of the decay period 
($t_{\rm{exp}}$ and $t_{\rm{exp}}+t_{\rm{dec}}$), weighted over all target isotopes, is written 
to the decay output file.

\section{Comparisons between measurements and calculations}
\label{sec:comparisons}

\begin{figure*}[!bht]
\begin{center}
\includegraphics[angle=-90,width=0.48\textwidth]{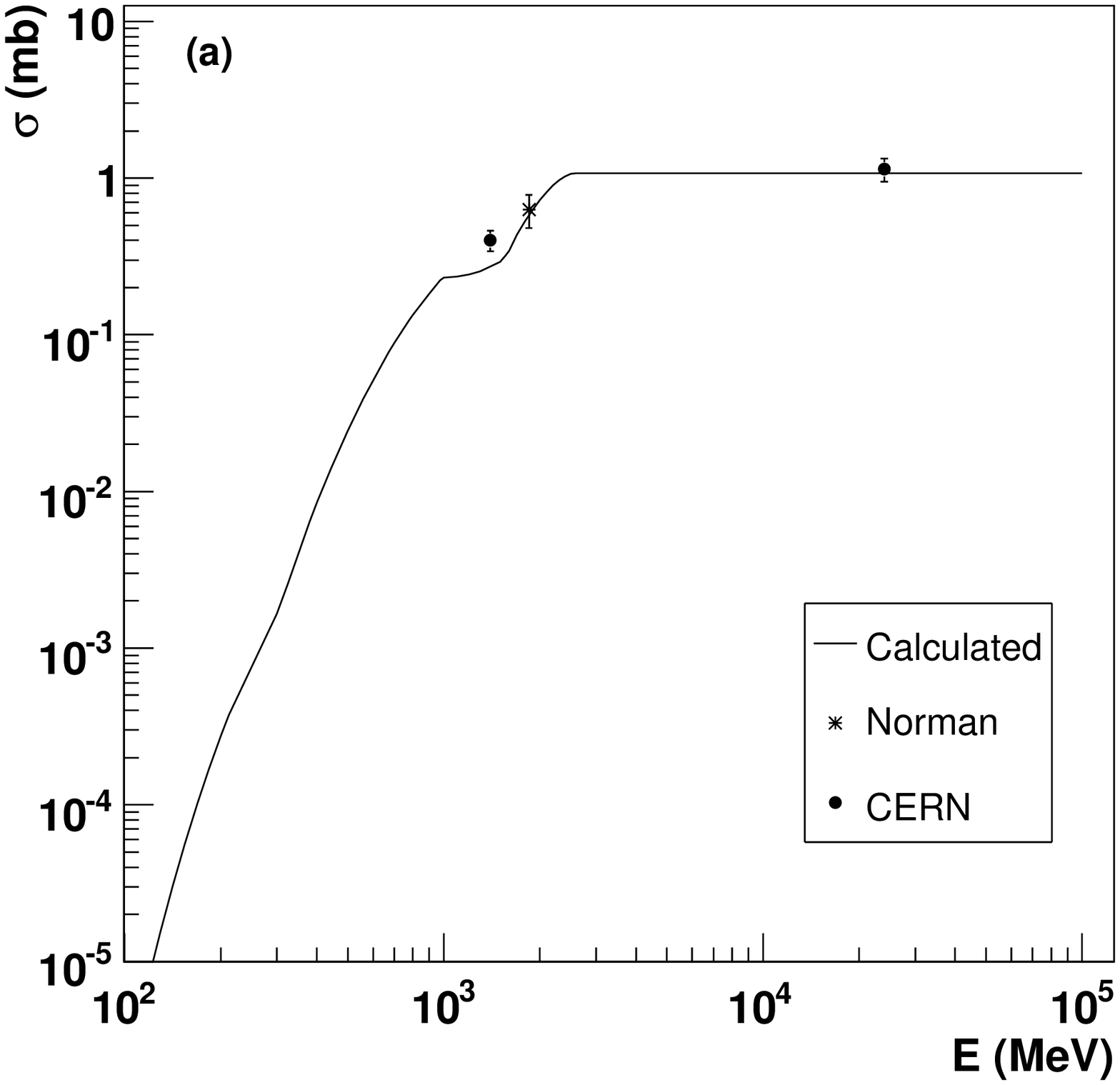}
\includegraphics[angle=-90,width=0.48\textwidth]{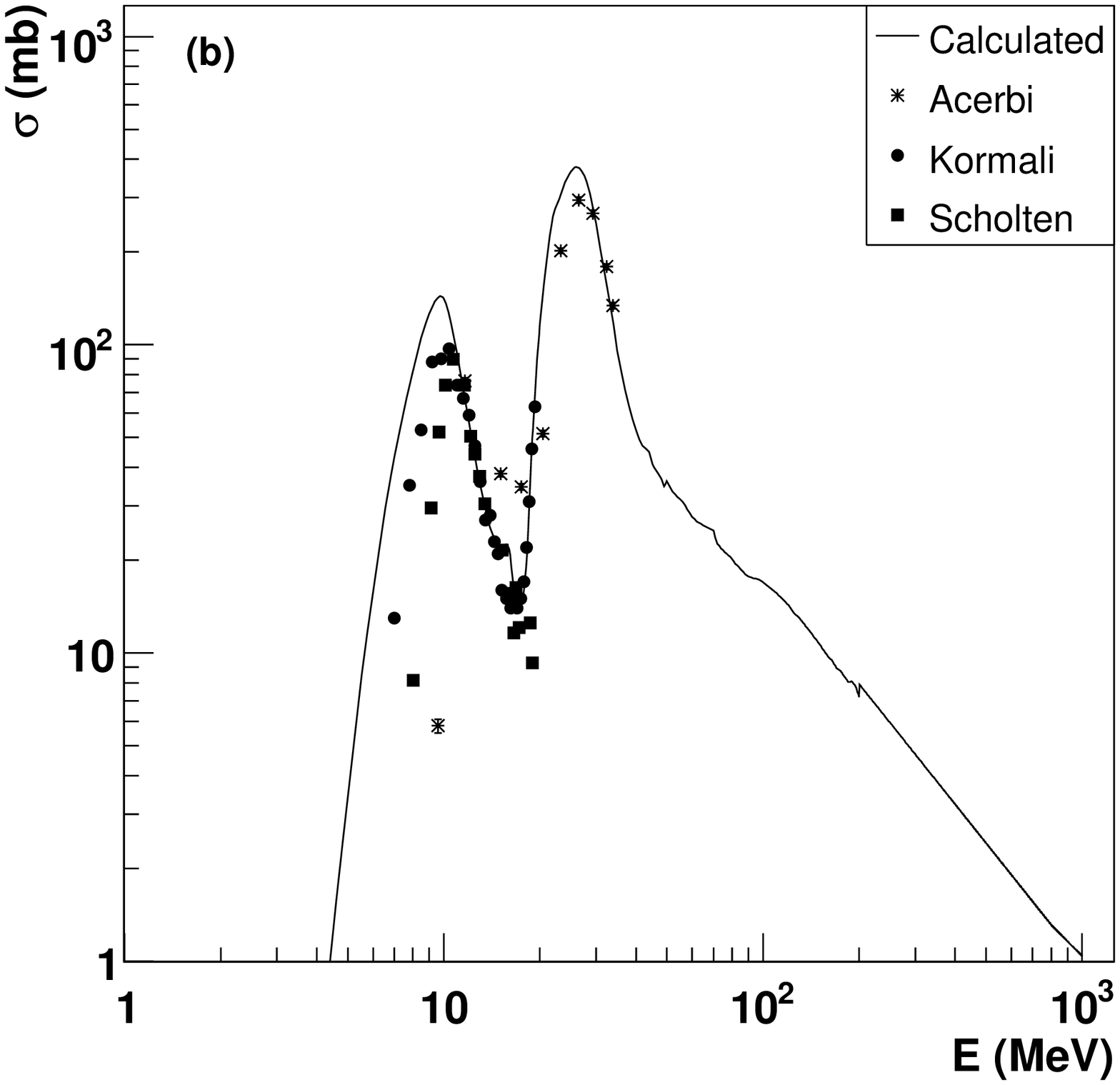}
\caption
{Cross-section as a function of energy for (a) $^{60}$Co and (b) $^{128}$I
produced by protons in natural tellurium. Data sets in (a) are from
Norman~\cite{Norman} and CERN~\cite{CERN}. Data points in (b) are from 
Acerbi~\cite{AcerbiI}, Kormali~\cite{KormaliI} and Scholten~\cite{ScholtenI}.}
\label{fig:NatTe_Co60}
\end{center}
\end{figure*}
\begin{figure*}[!htb]
\begin{center}
\includegraphics[angle=-90,width=0.48\textwidth]{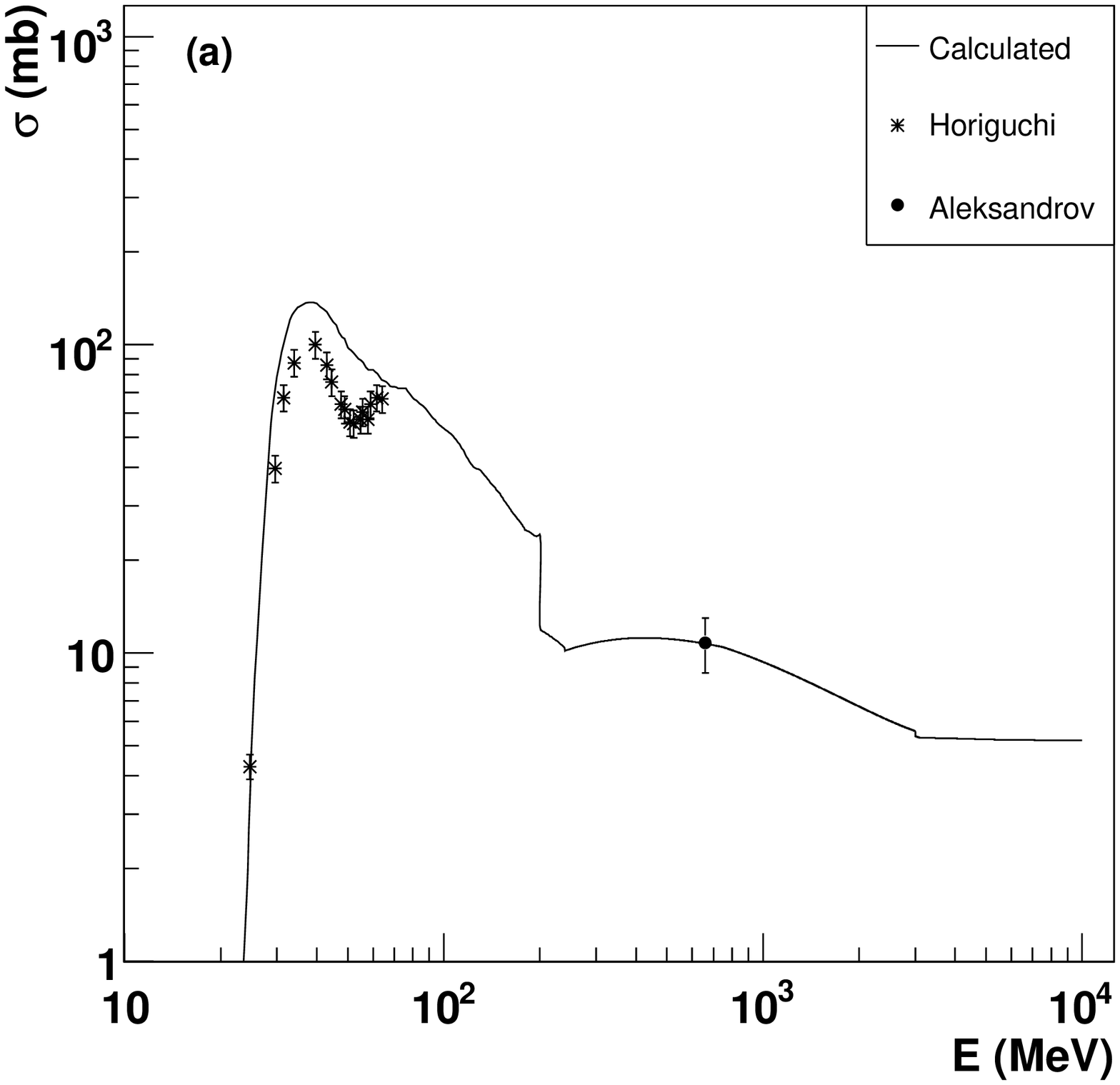}
\includegraphics[angle=-90,width=0.48\textwidth]{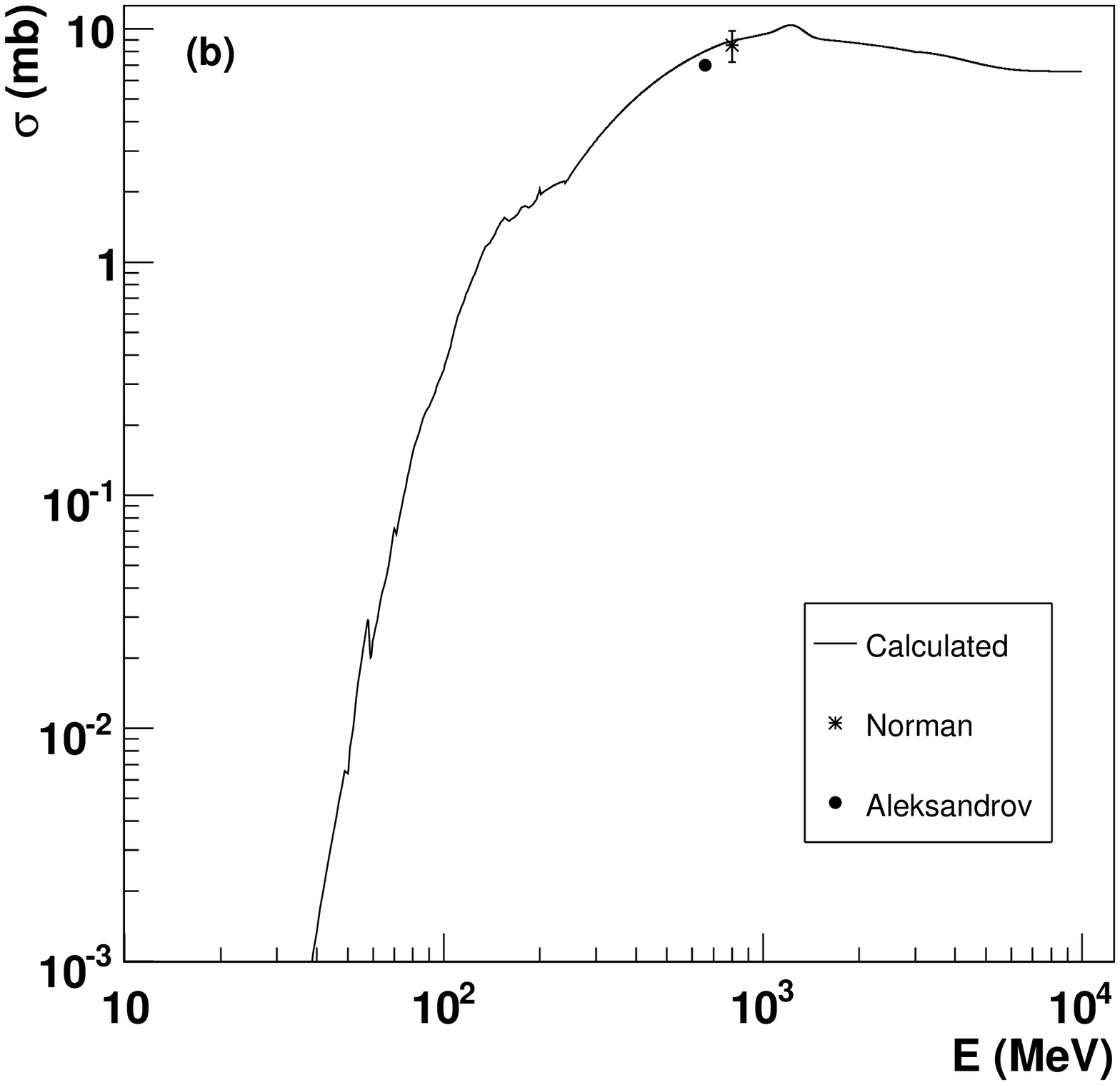}
\caption
{Cross-section as a function of energy for (a) $^{68}$Ge and (b) $^{60}$Co
produced by protons in natural germanium. Data points are from Norman~\cite{Norman},
Horiguchi~\cite{HoriguchiI} and Aleksandrov~\cite{Aleksandrov}.}
\label{fig:NatGe_Co60}
\end{center}
\end{figure*}
\begin{figure*}[!htb]
\begin{center}
\includegraphics[angle=-90,width=0.48\textwidth]{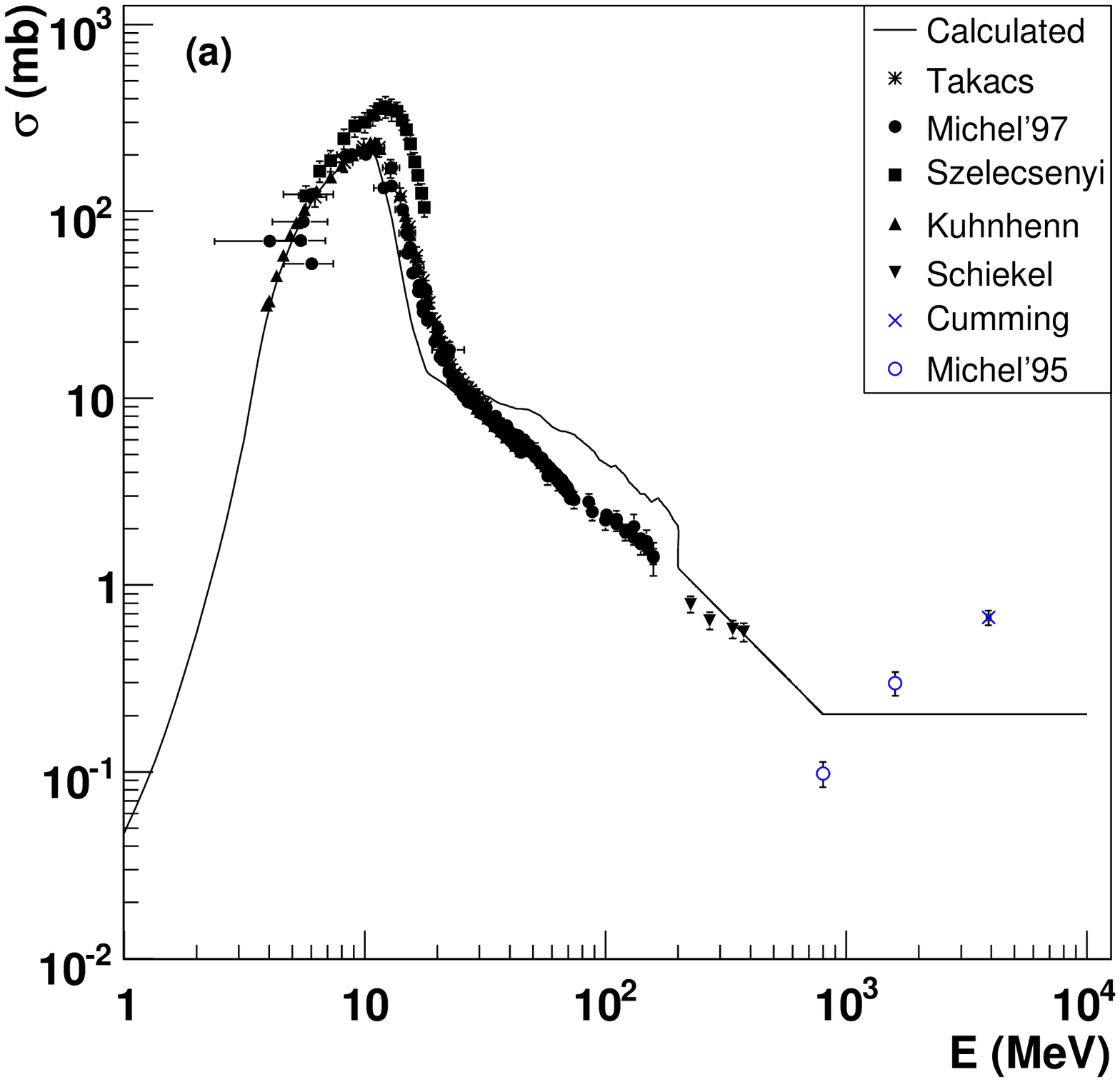}
\includegraphics[angle=-90,width=0.48\textwidth]{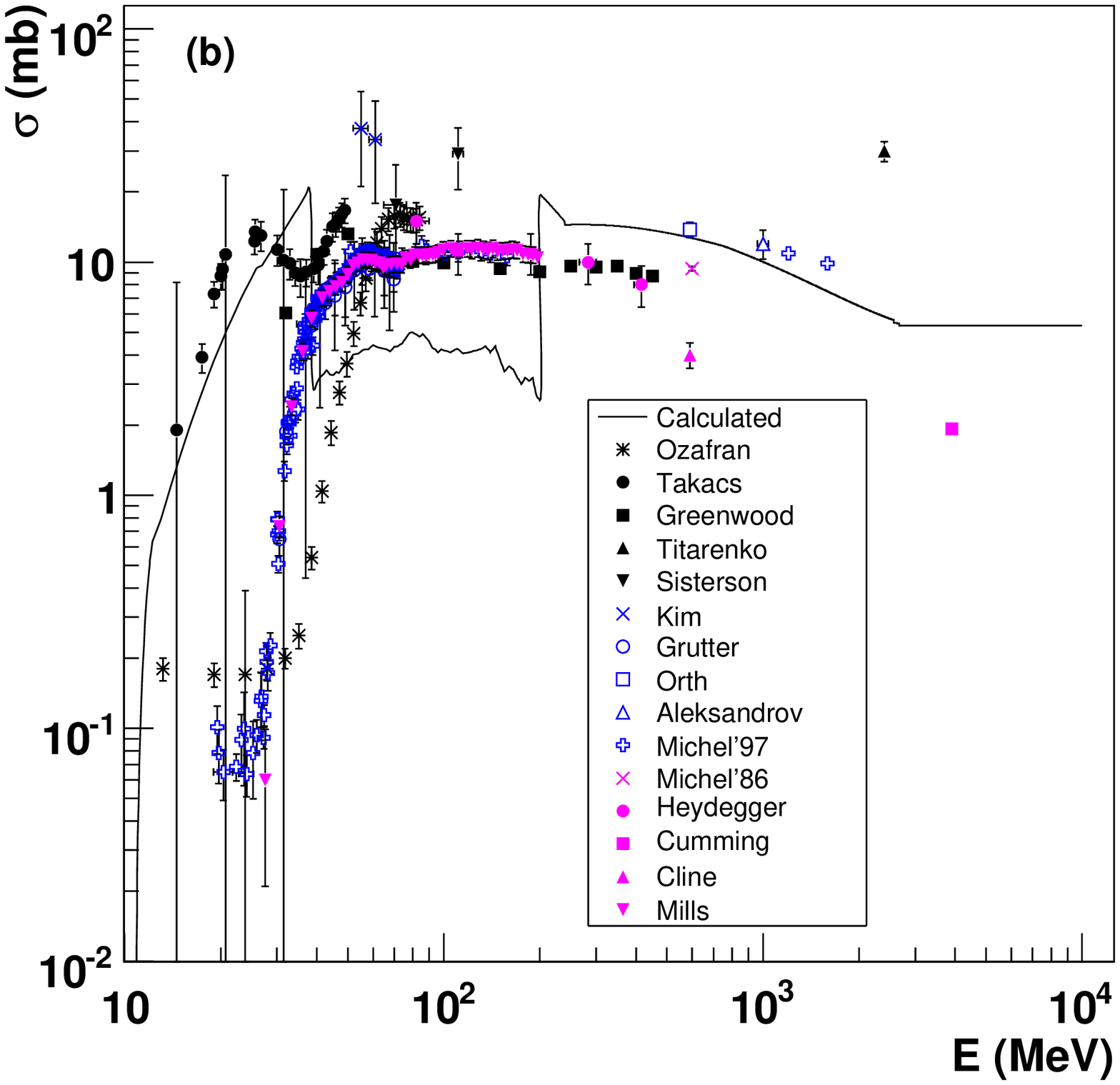}
\caption
{Cross-section as a function of energy for (a) $^{65}$Zn and (b) $^{60}$Co
produced by protons hitting a natural copper target. The calculation of the 
$^{60}$Co cross-section for energies in the range 40--200\,MeV uses the MENDL-2P data tables,
which underestimate the values obtained by various experiments.
Data points shown in (a) are from Tak\'{a}cs~\cite{Takacs}, Michel'97~\cite{Michel97}, 
Szelecs\'{e}nyi~\cite{Szelecsenyi}, Kuhnhenn~\cite{Kuhnhenn}, Schiekel~\cite{Schiekel},
Cumming~\cite{Cumming} and Michel'95~\cite{Michel95}. Data points shown in (b) are from
Ozafran~\cite{Ozafran}, Tak\'{a}cs~\cite{TakacsII}, Greenwood~\cite{Greenwood}, 
Titarenko~\cite{Titarenko}, Sisterson~\cite{Sisterson}, Kim~\cite{Kim}, Gr\"{u}tter~\cite{Grutter},
Orth~\cite{Orth}, Aleksandrov~\cite{AleksandrovII},
Michel'86~\cite{Michel86}, Heydegger~\cite{Heydegger},
Cline~\cite{Cline} and Mills~\cite{Mills}.}
\label{fig:NatCu_Co60}
\end{center}
\end{figure*}

The calculations of the cross-sections have been benchmarked using experimental data
from the EXFOR database~\cite{EXFOR} for natural tellurium, germanium and copper targets.
The results of the calculations agree in general with those shown in various reports from the IDEA
collaboration~\cite{IDEA}. Figures~\ref{fig:NatTe_Co60} to~\ref{fig:NatCu_Co60} show
the calculated cross-section graphs of several isotopes along with the data points from
experimental results~\cite{EXFOR}.
The agreement between calculations and experimental results is not uniform; some results
agree very well ($<$50\% relative difference), others agree within a factor of two, 
but there are also discrepancies up to a factor of 10 or more. 
It is clear that some of the MENDL-2P~\cite{MENDL2P}
cross-sections for energies between $\sim$10\,MeV to 200\,MeV do not agree very well 
with experimental results, such as those shown in Fig.~\ref{fig:NatCu_Co60}.
Obviously, the agreement will improve if the MENDL-2P data tables
are updated accordingly. In fact, the user is free to use other experimental
data for the cross-section tables instead of those from the MENDL-2P database.
Instructions on creating self-made data tables are provided in the README file of the
package~\cite{Code}.

\begin{table}[!hbt]
\caption{\label{tab:TeYields}Induced activities $Y_j^{\rm{dec}}$ ($\mu$Bq\,kg$^{-1}$) of TeO$_2$ for 
three nuclides assuming four months of exposure to cosmic rays and two years of decay in a 
shielded underground environment. There are two calculations from
the IDEA Collaboration~\cite{IDEA}; IDEA$^{\rm{a}}$ uses the cosmic-ray flux distribution $\phi$ from
Fig.~\ref{fig:cosrayspectrum} while IDEA$^{\rm{b}}$ uses the distribution from Ziegler~\cite{Ziegler}.\newline}
\begin{tabular*}{0.475\textwidth}{@{\extracolsep{\fill}}llll}
\hline
         & $^{60}$Co & $^{125}$Sb &  $^{124}$Sb \\
\hline
CUORE~\cite{CUORE}   & 0.20       &  15          &  0.05 \\
IDEA$^{\rm{a}}$~\cite{IDEA} & 0.02       &  $6 \pm 6$   &  $0.01 \pm 0.03$ \\
IDEA$^{\rm{b}}$~\cite{IDEA} & 0.02       &  $22 \pm 15$ &  $0.05 \pm 0.06$ \\
ACTIVIA              & 0.03       &  6           &  0.02 \\
\hline
\end{tabular*}
\end{table}
\begin{table*}[!htb]
\caption{\label{tab:GeYields1}Induced rates of isotope production $Y_j$ (kg$^{-1}$ day$^{-1}$) 
in natural germanium at sea level. The values from Avignone~\cite{Avignone} are separated into 
Monte Carlo simulation (MC) and experimental (exp) results.}
\begin{tabular*}{0.975\textwidth}{@{\extracolsep{\fill}}llllllllll}
\hline
 & $^{68}$Ge & $^{60}$Co  & $^{65}$Zn & $^{58}$Co & $^{57}$Co & $^{56}$Co & $^{54}$Mn & $^{63}$Ni & $^{55}$Fe \\
\hline
HMS-ALICE,YIELDX~\cite{HMSALICE,YIELDX}  & 89.0 &  4.8 & 77.0 & 14.0 &  9.7 &  3.0 &  7.2 &  5.2 &  8.0 \\
Avignone~\cite{Avignone} (MC)  & 29.6 &  --- & 34.4 &  5.3 &  4.4 &  --- &  2.7 &  --- & ---  \\
Avignone~\cite{Avignone} (exp) & $30.0 \pm 7.0$ &  --- & $38.0 \pm 6.0$ & $3.5 \pm 0.9$ & $2.9 \pm 0.4$ &  --- & $3.3 \pm 0.8$ &  --- & ---  \\
Klapdor-Kleingrothaus~\cite{Klapdor}     & 58.4 &  6.6 & 79.0 & 16.1 & 10.2 &  --- &  9.1 &  4.6 &  8.4 \\
Miley~\cite{Miley}         & 26.5 &  4.8 & 30.0 &  4.4 &  0.5 &  --- &  --- &  --- &  --- \\
Barabanov~\cite{Barabanov} & 81.6 &  2.9 &  --- &  --- &  --- &  --- &  --- &  --- &  --- \\
Formaggio~\cite{Formaggio} & 29.5 &  --- &  --- & 154.0 & --- &  5.4 &  4.8 &  --- &  --- \\
ACTIVIA                    & 45.8 &  2.8 & 29.0 &  8.5 &  6.7 &  2.0 &  2.7 &  1.6 &  3.4 \\
\hline
\end{tabular*}
\end{table*}
\begin{table*}[!htp]
\caption{\label{tab:GeYields2}Induced rates of isotope production 
$Y_j$ (kg$^{-1}$ day$^{-1}$) in enriched germanium 
(86\% $^{76}$Ge and 14\% $^{74}$Ge) at sea level.}
\begin{tabular*}{0.975\textwidth}{@{\extracolsep{\fill}}lllllllll}
\hline
 & $^{68}$Ge & $^{60}$Co & $^{65}$Zn & $^{58}$Co & $^{57}$Co & $^{54}$Mn & $^{63}$Ni & $^{55}$Fe \\
\hline
HMS-ALICE,YIELDX~\cite{HMSALICE,YIELDX} & 13.0 &  6.7 & 24.0 &  6.2 &  2.3 &  5.4 &  6.0 &  2.3 \\
Avignone~\cite{Avignone}   &  0.9 &  --- &  6.4 &  1.8 &  1.0 &  1.4 &  --- & --- \\
Miley~\cite{Miley}         &  1.2 &  3.5 &  6.0 &  1.6 &  0.1 &  --- &  --- & --- \\
Barabanov~\cite{Barabanov} &  5.8 &  3.3 &  --- &  --- &  --- &  --- &  --- & --- \\
ACTIVIA                    &  7.6 &  2.4 & 10.4 &  5.5 &  2.9 &  2.2 &  1.4 & 1.6 \\
\hline
\end{tabular*}
\end{table*}
\begin{table*}[!tbp]
\caption{\label{tab:CuYields}Induced rates of isotope production $Y_j$ (kg$^{-1}$ day$^{-1}$) in 
natural copper at sea level.
For the ACTIVIA calculations, there are two sets of values; those from only the semi-empirical
formulae and those that also include the MENDL-2P data.}
\begin{tabular*}{0.975\textwidth}{@{\extracolsep{\fill}}llllllll}
\hline
 & $^{56}$Co & $^{57}$ Co & $^{58}$Co & $^{60}$Co & $^{54}$Mn & $^{59}$Fe & $^{46}$Sc \\
\hline
IDEA~\cite{IDEA}           & 22.9  & 88.3 & 159.6 & 97.4 & 32.5  &  6.5 &  3.8 \\
Baudis~\cite{Baudis}       &  ---  & 30.5 &   --- & 25.7 & 134.2 &  --- &  --- \\
Formaggio~\cite{Formaggio} & 972.3 &  --- & 131.5 & 2675.5 & 52.2 & --- &  --- \\
ACTIVIA formulae only      &  8.7  & 32.5 & 56.6  & 26.3   & 14.3 &  4.2 &  3.1 \\
ACTIVIA with MENDL-2P      & 14.1 & 36.4 & 38.1 & 9.7 & 12.5 & 1.8 & 3.1 \\
\hline
\end{tabular*}
\end{table*}

Comparisons have been made between the induced activity calculated by ACTIVIA
with results shown in reports from the IDEA collaboration~\cite{IDEA} 
for several targets and isotope products.
Table~\ref{tab:TeYields} shows an example of the calculated induced activities for TeO$_2$
exposed to cosmic rays at sea level for four months and shielded underground for a further
two years. Note that the ACTIVIA result for $^{60}$Co, which
predicts a production rate $Y_j = 0.8\mu$Bq\,kg$^{-1}$ 
(YIELDX~\cite{YIELDX} predicts $0.7\mu$Bq\,kg$^{-1}$), is a factor
of 10 lower than the estimate obtained by the CUORE collaboration~\cite{CUORE} (using a
modified version of COSMO~\cite{ModCosmo})
owing to the improved calculation of the cross-section (see Fig.~\ref{fig:NatTe_Co60}).
Tables~\ref{tab:GeYields1} and~\ref{tab:GeYields2} show the induced rates of isotope production 
by cosmic rays at sea level for natural and enriched germanium, respectively.
One can see that there are differences between the various estimates presented in the tables.
For example, the ACTIVIA results using the MENDL-2P data tables shown 
in Table~\ref{tab:GeYields1} are generally quite close to those 
from Avignone~\cite{Avignone}, but are approximately a factor of two less than those from
HMS-ALICE~\cite{HMSALICE} and YIELDX~\cite{YIELDX}
estimates and those from Klapdor-Kleingrothaus~\cite{Klapdor}.
Similar discrepancies between measurements and calculations are also found for the enriched 
germanium yield estimates shown in Table~\ref{tab:GeYields2}.
Table~\ref{tab:CuYields} shows the isotope production yields for natural copper.
Two of the ACTIVIA results agree with those from Baudis and Schnee~\cite{Baudis}, but all of the 
calculated yields are less than those estimated by the IDEA group, whose calculations
included the MENDL low energy ($E \lesssim 10$\,MeV) neutron data~\cite{MENDL2}.
Most of the cobalt results from Formaggio~\cite{Formaggio} in Tables~\ref{tab:GeYields1} 
and~\ref{tab:CuYields} are systematically higher than those from
other analyses; it is unclear why this is the case.

One important source of uncertainty in the production rates is the cosmic ray spectrum, where
discrepancies around a factor of two can be explained by 
differences between the parameterisations used in various calculations 
(such as the spectrum shown in Fig.~\ref{fig:cosrayspectrum} used in ACTIVIA) and 
cosmic ray fluxes (which depend on geographic location and time) measured by various experiments.

Another important issue is that the semi-empirical formulae must be able to predict
the cross-sections of isotope production over a wide range of isotope masses and input beam energies. As
shown in Fig.~\ref{fig:STRegions}, there are several physical processes that can take place in nuclear
reactions. Each process must be described by a set of formulae that can adequately
predict the cross-section for any given isotope and beam energy. Obviously, this can only be
accomplished if there is enough experimental data to derive such functions and obtain
all relevant parameters with sufficient accuracy. The data used by Silberberg and Tsao did not cover
all possible isotopes and energies, which limits the predictive power of the formulae for various
cases. As can be seen in Figs.~\ref{fig:NatTe_Co60} 
and~\ref{fig:NatGe_Co60}, some isotopes have very few measurements available.
More experimental data are needed so that improvements can be made to semi-empirical formulae,
data tables and Monte Carlo simulations for isotope production calculations.

\section{Code modifications and extensions}
\label{sec:extensions}

The code is structured so that modifications or extensions can be made
to several components of the cross-section and yield calculations whilst keeping
the overall framework intact.

Different input beam spectra can be used by including classes that inherit from
ActBeamSpectrum; the appropriate spectrum must be chosen by the input class. For example,
a class describing the flux of an ion beam can be provided, although the formulae for calculating
the cross-sections of products from such a beam would need to be implemented as well.
The beam spectrum class also needs to specify the energy range and bin width
that are used to calculate the integral in Eq.~\ref{eqn:yield1}, as well as the 
atomic number $Z_{\rm{beam}}$ and mass number $A_{\rm{beam}}$ of the beam particles.

The ``decayData.dat'' file contains a list of all available isotope products with half-lives greater
than one day, as well as their side-branch nuclei. Other isotopes, with shorter half-lives,
can be added to this data file. The code will calculate the cross-sections and yields 
for these nuclei automatically without any changes necessary to the source code. However,
any side-branches matching the new isotopes would need to be removed to avoid double counting.

At present, the cross-sections are calculated by the
ActSTXSecAlgorithm class using a combination of values from (MENDL) 
data tables as well as semi-empirical formulae from Silberberg and Tsao. 
Cross-section models for processes like spallation and fission are implemented in 
their own classes, making it easier to update their parameters or formulae when
necessary. New nuclear models can be added to the ActSTXSecAlgorithm class if required.

Other cross-section algorithms can be specified, as long as
they inherit from the base class ActAbsXSecAlgorithm and are selected appropriately 
in the input class. For example, new algorithms may be needed to specify cross-section 
formulae when beams other than cosmic rays (protons/neutrons) are used.

Other decay algorithms could be written, such as those that take into account
chained decays, if they follow the ActAbsDecayAlgorithm abstract interface,
which has access to the list of possible isotope products (ActProdNuclideList),
as well as the definition of the target (ActTarget) and the ActProdXSecData results it stores 
(the target-product cross-sections).
ActNuclide objects can be retrieved from the target and product list to take
care of any bookkeeping required for chained decay algorithms.
Note that the decay algorithm is separated from the cross-section calculations,
allowing the possibility of using different decay algorithms for the same initial production yields,
provided they are selected appropriately by the input class.

Changes can be made to how the calculations are set up. The default ActInput class uses
standard \Cpp\ input/output commands to get the necessary information from the user.
Other classes using different input formats, such as XML files, 
could be implemented. However,
they must all derive from the abstract ActAbsInput interface, and provide all required
inputs (beam spectrum, target and product isotopes, decay times and selection of
cross-section and decay yield algorithms).

At present there are two output formats available: ASCII text files and ROOT~\cite{ROOT} files.
Other output formats could be written, provided the classes follow the abstract ActAbsOutput
interface and know how to deal with lines of text, tables of data (ActOutputTables)
and graphs (ActAbsGraph objects).

\section{Summary}
\label{sec:Summary}
We have presented the ACTIVIA \Cpp\ computer package that can be used to calculate
target-product cross-sections and the production and decay yields of isotopes from 
cosmic ray activation. Cross-sections are calculated using a combination of semi-empirical
formulae, assumed to be the same for proton and neutron beams, 
and ASCII tables of (experimental) data.

The basic structure of the code, designed to work in the Unix
environment, has been presented. It is structured so that modifications or extensions can be made
to several components of the cross-section and yield calculations whilst keeping
the overall framework intact. Since ACTIVIA uses only formulae or data tables to calculate
cross-sections, and does not simulate individual particles and their interactions,
it is easier to use than full Monte Carlo simulation
computer packages such as GEANT4~\cite{Geant4}, while it is more extensible than similar
FORTRAN-based codes like COSMO~\cite{ModCosmo}. 
At present, the code is limited to only calculating the
isotope production cross-sections and yields for one given target element
(with its various isotopes); for targets made up of
several elements, the code must be run separately for each and the results 
combined at the end by the user. Results from the calculations can be written out
as ASCII text or stored within ROOT~\cite{ROOT} files.

We have compared isotope production results from ACTIVIA against experimental data and 
estimates from other calculations; in general, they agree within a factor of two. 
It is possible to change or extend the algorithms and formulae used to calculate the 
cross-sections and isotope yields. 
Finally, we have discussed ways of modifying the code such that it may be used
in applications other than calculating yields of cosmogenic isotope products.

\section*{Acknowledgments}
\label{sec:Acknowledgments}

We thank Tom Latham and Ben Morgan for useful discussions about the code design.
We also thank the OECD Nuclear Energy Agency, France, for providing us with the MENDL-2P
data tables. This work is supported by the Science and Technology Facilities Council 
(United Kingdom).




\begin{thebibliography}{99}

\bibitem{ST1}
R. Silberberg, C.H. Tsao, Astrophys. J. Suppl. 220 (I)  25 (1973) 315;
R. Silberberg, C.H. Tsao, Astrophys. J. Suppl. 220 (II)  25 (1973) 335.

\bibitem{ST2}
R. Silberberg, C.H. Tsao, Astrophys. J. Suppl.  35 (1977) 129;
R. Silberberg, C.H. Tsao, J.R. Letaw, Astrophys. J. Suppl.  58 (1985) 873;
R. Silberberg, C.H. Tsao, J.R. Letaw, in: Proceedings of the 20th 
International Cosmic Ray Conference, Moscow, vol. 2, 1987, p. 133;
R. Silberberg, C.H. Tsao, Phys. Rep. 191 (1990) 351;
R. Silberberg, C.H. Tsao, A.F. Barghouty, Astrophys. J. 501 (1998) 911.

\bibitem{tritium}
A.Yu. Konobeyev, Yu.A. Korovin, Nucl. Instr. and Meth. B 82 (1993) 103.

\bibitem{EXFOR}
The Experimental Nuclear Reaction Data (EXFOR) database at 
\url{http://www-nds.iaea.org/exfor/}

\bibitem{MENDL2P}
Yu.N. Shubin, \etal, MENDL-2P: Proton Reaction Data Library for 
Nuclear Activation (Medium Energy Nuclear Data Library) IAEA-NDS-204, 1998.

\bibitem{Armstrong}
T.W. Armstrong, K.C. Chandler, J. Barish, J. Geophys. Res. 78 (1973) 2715.

\bibitem{Gehrels}
N. Gehrels, Nucl. Instr. and Meth. A 239 (1985) 324.

\bibitem{Code}
The ACTIVIA package is available for download at
\url{http://www.warwick.ac.uk/go/activia}

\bibitem{Seeger}
P.A. Seeger, Nucl. Phys. 25 (1961) 1.

\bibitem{ROOT}
R. Brun, F. Rademakers, Nucl. Instr. and Meth. A 389
(1997) 81. See also \url{http://root.cern.ch}

\bibitem{IDEA}
Integrated double-beta decay European activities at
\url{http://idea.dipscfm.uninsubria.it}

\bibitem{Norman}
E.B. Norman, \etal, Nucl. Phys. B Proc. Suppl. 143 (2005) 508.

\bibitem{CERN}
Unpublished preliminary results from irradiation experiments at CERN.

\bibitem{AcerbiI}
E. Acerbi, \etal, Int. J. Appl. Radiat. Isot. 26 (1975) 741.

\bibitem{KormaliI}
S.M. Kormali, D.L. Swindle, E.A. Schweikert, J. Radioanal. Nucl. Chem. 31 (1976) 437.

\bibitem{ScholtenI}
B. Scholten, S.M. Qaim, G. St\"{o}cklin, Int. J. Appl. Radiat. Isot. 40 (1989) 127.

\bibitem{HoriguchiI}
T. Horiguchi, \etal, Int. J. Appl. Radiat. Isot. 34 (1983) 1531.

\bibitem{Aleksandrov}
Yu.V. Aleksandrov, \etal, Bull. Russ. Acad. Sci. Phys. 59 (1995) 895.

\bibitem{Takacs}
S. Tak\'{a}cs, \etal, Nucl. Instr. and Meth. B 188 (2002) 106.

\bibitem{Michel97}
R. Michel, \etal, Nucl. Instr. and Meth. B 129 (1997) 153.

\bibitem{Szelecsenyi}
F. Szelecs\'{e}nyi, \etal, Nucl. Instr. and Meth. B 174 (2001) 47.

\bibitem{Kuhnhenn}
J. Kuhnhenn, Ph.D. Thesis, University of K\"{o}ln, Germany, 2001.

\bibitem{Schiekel}
Th. Schiekel, \etal, Nucl. Instr. and Meth. B 114 (1996) 91.

\bibitem{Cumming}
J. B. Cumming, \etal, Phys. Rev. C 10 (1974) 739.

\bibitem{Michel95}
R. Michel, \etal, Nucl. Instr. and Meth. B 103 (1995) 183.

\bibitem{Ozafran}
M.J. Ozafran, \etal, J. Radioanal. Nucl. Chem. 131 (1989) 467.

\bibitem{TakacsII}
S. Tak\'{a}cs, \etal, Nucl. Instr. and Meth. B 251 (2006) 56.

\bibitem{Greenwood}
L.R. Greenwood, R.K. Smither, Prog. Rep. U.S. Department of Energy, 
Fusion Energy Series, No. 0046, 18 (1984) 11.

\bibitem{Titarenko}
Yu.E. Titarenko, \etal, Proc. AccApp'03, San Diego, California,
nucl-ex/0305026 (2003).

\bibitem{Sisterson}
J.M. Sisterson, \etal, Nucl. Instr. and Meth. B 240 (2005) 617.

\bibitem{Kim}
E.J. Kim, \etal, J. Nucl. Sci. Tech. 36 (1999) 29.

\bibitem{Grutter}
A. Gr\"{u}tter, Nucl. Phys. A 383 (1982) 98.

\bibitem{Orth}
C.J. Orth, \etal, J. Inorg. Nucl. Chem. 38 (1976) 13.

\bibitem{AleksandrovII}
Yu.V. Aleksandrov, \etal, Bull. Russ. Acad. Sci. Phys. 54 (1990) 161.

\bibitem{Michel86}
R. Michel, \etal, Nucl. Instr. and Meth. B 16 (1986) 61.

\bibitem{Heydegger}
H.R. Heydegger, C.K. Garrett, A. Van Ginneken, Phys. Rev. C 6 (1972) 1235.

\bibitem{Cline}
J.E. Cline, E.B. Nieschmidt, Nucl. Phys. A 169 (1971) 437.

\bibitem{Mills}
S.J. Mills, G.F. Steyn, F.M. Nortier, Int. J. Appl. Radiat. Isot. 43 (1992) 1019.


\bibitem{Ziegler}
J.F. Ziegler, IBM J. Res. Dev. 42 (1998) 117.

\bibitem{CUORE}
R. Ardito, \etal, hep-ex/0501010 (2005).

\bibitem{YIELDX}
Code written by Silberberg and Tsao implementing their own semi-empirical 
formulae described in Refs.~\cite{ST1,ST2}.

\bibitem{ModCosmo}
C.J. Martoff, P.D. Lewin, Comput. Phys. Commun. 72 (1992) 96.
Code modified by Y.A. Ramachers.

\bibitem{Avignone}
F.T. Avignone, \etal, Nucl. Phys. B Proc. Suppl. 28 (1) (1992) 280.

\bibitem{HMSALICE}
M. Blann, Phys. Rev. C 54 (1996) 1341;
M. Blann, M.B. Chadwick, Phys. Rev. C 57 (1998) 233.

\bibitem{Klapdor}
H.V. Klapdor-Kleingrothaus, \etal, Nucl. Instr. and Meth. A 481 (2002) 149.

\bibitem{Miley}
H.S. Miley, \etal, Nucl. Phys. B Proc. Suppl. 28 (I) (1992) 212.

\bibitem{Barabanov}
I. Barabanov, \etal, Nucl. Instr. and Meth. B 251 (2006) 115.

\bibitem{Formaggio}
J.A. Formaggio, C.J. Martoff, Ann. Rev. Nucl. Part. Sci. 54 (2004) 361.

\bibitem{Baudis}
L. Baudis, R.W. Schnee, Dark Matter Experiments NUSL White Paper, 2002.

\bibitem{MENDL2}
Yu.N. Shubin, \etal, MENDL-2: Neutron Reaction Data Library for Nuclear Activation
(Medium Energy Nuclear Data Library) IAEA-NDS-136, 1995.

\bibitem{Geant4}
S. Agostinelli, \etal, Nucl. Instr. and Meth. A 506 (2003) 250.

\end{thebibliography}
\end{document}